\documentclass{ws-procs975x65}

\begin{document}



\title{COMPACT BINARY INSPIRAL AND THE SCIENCE POTENTIAL OF THIRD-GENERATION GROUND-BASED GRAVITATIONAL WAVE DETECTORS\footnote{This research was supported in part by
PPARC grant PP/B500731/1.}}

\author{CHRIS VAN DEN BROECK and ANAND S.~SENGUPTA}

\address{School of Physics and Astronomy,\\
Cardiff University, 
Queen's Buildings, The Parade, 
Cardiff CF24 3AA\\
United Kingdom\\
\email{Chris.van-den-Broeck@astro.cf.ac.uk, Anand.Sengupta@astro.cf.ac.uk}}


\begin{abstract}
We consider EGO as a possible third-generation ground-based gravitational wave detector and evaluate its capabilities for the detection and interpretation of compact binary inspiral signals. We identify areas of astrophysics and cosmology where EGO would have qualitative advantages, using Advanced LIGO as a benchmark for comparison.
\end{abstract}
\bodymatter

\vskip 0.1cm

\emph{Compact binary inspiral.}
Inspirals of compact binary objects (black holes and/or neutron stars) are among the most promising sources for ground-based gravitational wave detectors \cite{Grishchuk}, and as such they are eminently suitable to evaluate the science potential of future observatories. In the quasi-circular, adiabatic regime, where the periods of orbits are much smaller than the inspiral timescale, gravitational waveforms have been computed in the post-Newtonian (PN) approximation,\cite{Blanchet} where the signal is a superposition of harmonics in the orbital phase. Recently the full waveforms, with inclusion of PN amplitude corrections, were used to accurately assess the potential of Advanced LIGO and EGO in terms of redshift reach, detection rates, and parameter estimation.\cite{amppapers} Here we briefly discuss possible implications for astrophysics and cosmology; for the theoretical underpinnings as well as complete references we refer to these more technical papers.

\begin{figure}[htbp!]
\begin{center}
\epsfig{file=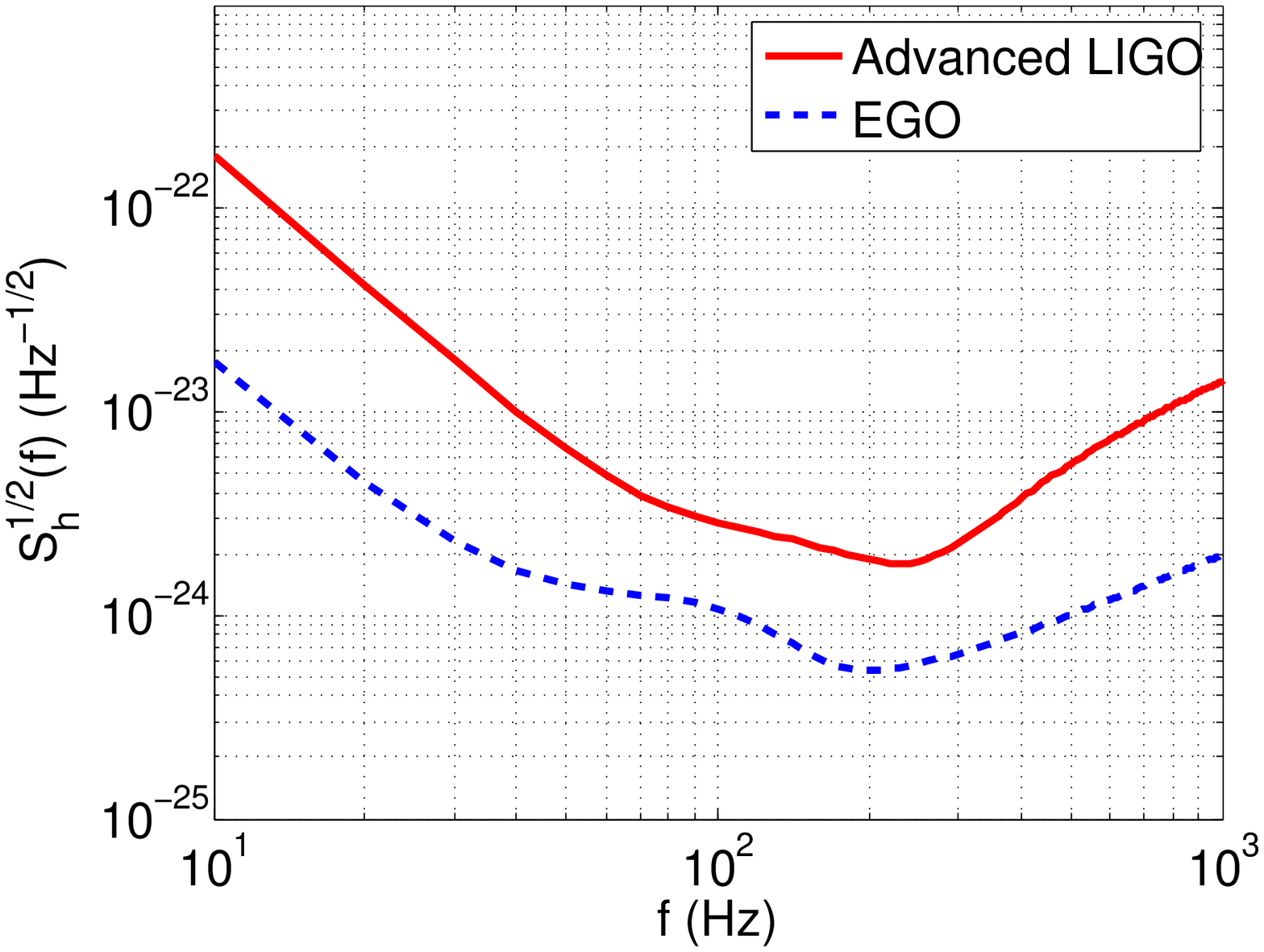,height=4.5cm,width=5.5cm,angle=0}
\epsfig{file = 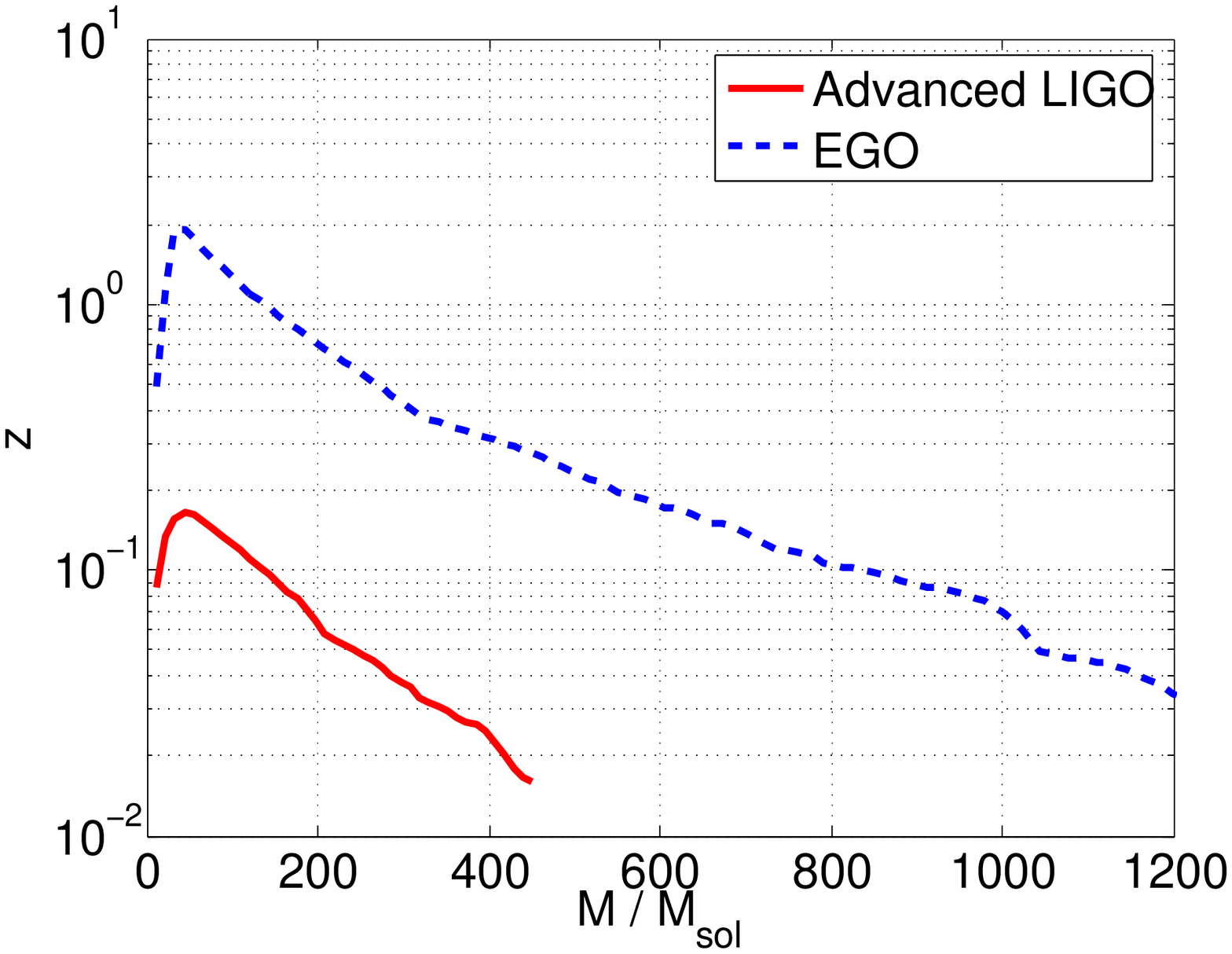,height=4.5cm,width=5.5cm,angle=0}
\end{center}
\caption{Plots of the stain sensitivities of EGO and Advanced LIGO (left) and their redshift reach for a fixed SNR of 10 (right). On the right hand side we have fixed $m_1/m_2 = 0.1$ and angles $\theta=\phi=\pi/6$, $\psi=\pi/4$, $\iota=\pi/3$.}
\label{f:strain_z}
\end{figure}

\emph{EGO as a third-generation detector.}
EGO is not yet on the drawing boards; rather, its strain sensitivity as plotted in Fig.~\ref{f:strain_z} should be viewed as a summary of what is believed to be possible with steadfast advances in interferometer technology over the next decade or so. In most of the frequency interval shown, the difference in sensitivity between EGO and Advanced LIGO is a factor of a few; at low frequencies, which are of interest for compact binary inspiral, the difference is about an order of magnitude.

\emph{Redshift and mass reach.} The right hand panel of Fig.~\ref{f:strain_z} shows how these sensitivities translate into redshift reach as a function of total mass $M$ for a fixed ratio of the component masses $m_1$, $m_2$. The mass reach of Advanced LIGO is slightly over $400\,M_\odot$ while EGO can see systems that are three times heavier. It is useful to make a distinction between stellar mass systems with $M \lesssim 100\,M_\odot$ and the heavier intermediate mass binaries with $M$ up to $\mbox{(a few)}\,\times\,1000\,M_\odot$. The latter systems may form in the centers of galaxies and in globular clusters, and they are expected to be rather asymmetric; hence our choice $m_1/m_2 = 0.1$. (Note that the redshift reach would be larger in the equal mass case, and for a more convenient sky position and orientation.) We see that EGO would be able to detect stellar mass inspirals through much of the visible Universe. Detection rates in EGO have been conservatively estimated to be at least $700$ times higher than in Advanced LIGO.\cite{amppapers} 

\emph{Measuring component masses.} How well can parameters be extracted from a signal in EGO? The individual component masses $m_1$, $m_2$ enter the waveforms through particular combinations, the chirp mass ${\cal M}$ and the symmetric mass ratio $\eta$. As a consequence, the latter tend to be measurable with good accuracy, while $m_1$, $m_2$, which are of direct astrophysical interest, generally are not well-determined in initial detectors.
\begin{figure}[htbp!]
\begin{center}
\epsfig{file=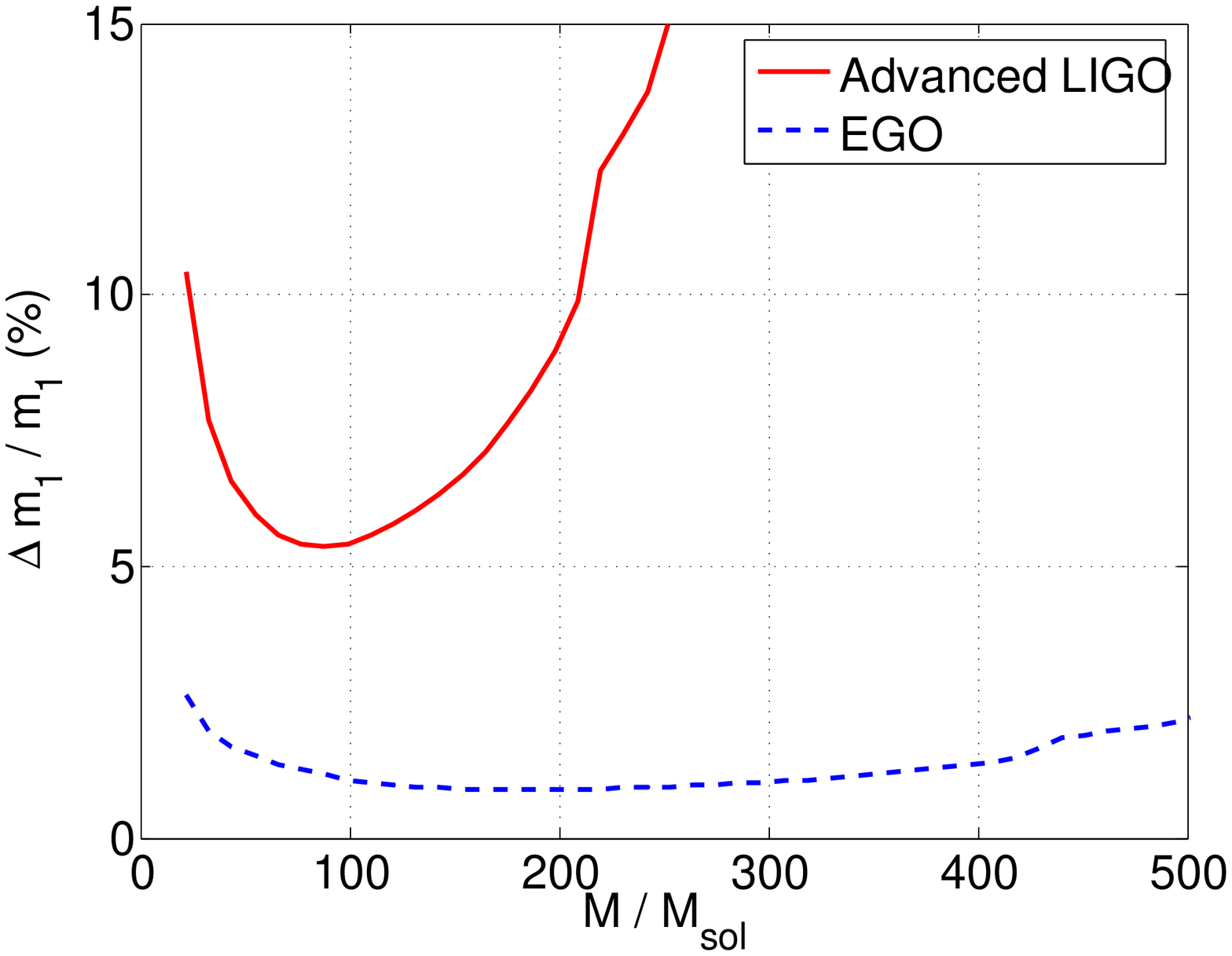,height=4.5cm,width=5.5cm,angle=0}
\epsfig{file=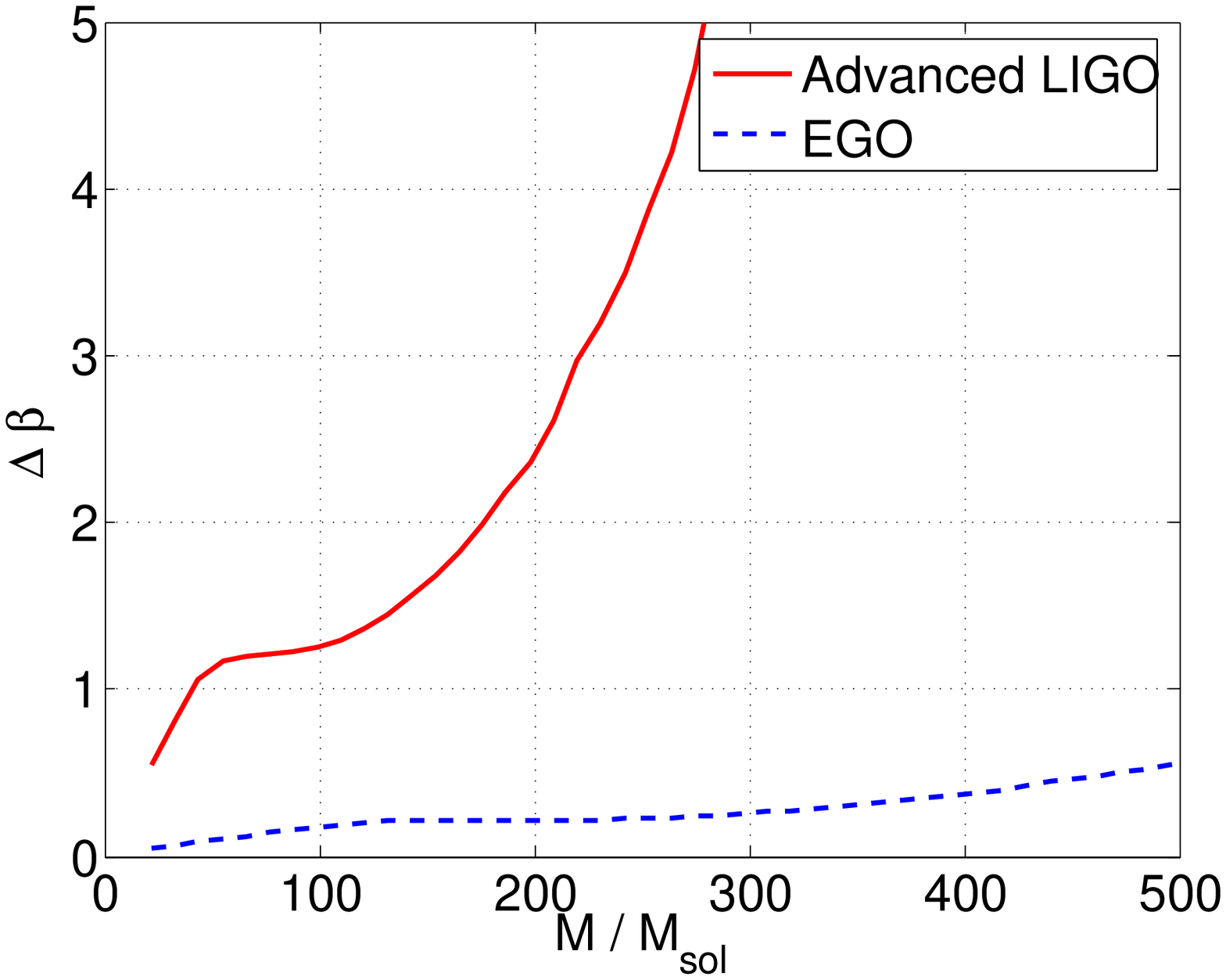,height=4.5cm,width=5.5cm,angle=0}
\end{center}
\caption{Relative error on component mass (left) and error on the spin-orbit parameter (right) at a distance of 100 Mpc, again setting $m_1/m_2=0.1$.}
\label{f:compmasses_beta}
\end{figure}  
In the left panel of Fig.~\ref{f:compmasses_beta} we see that for a distance of 100 Mpc, in Advanced LIGO the relative error on component mass $m_1$ never goes below 5\%, while in EGO it is only a few percent in a very large mass range. EGO would enable us to ``map'' the mass distribution of black holes. It would give us a direct view on the way intermediate mass black holes grow through successive coalescences with smaller compact objects. Fig.~\ref{f:strain_z} indicates that for stellar mass systems, parameter estimation in EGO up to redshift $z \sim 2$ (corresponding to a luminosity distance $\sim$ 16 Gpc) would be as good as in Advanced LIGO up to only $z \sim 0.2$ (or $\sim$ 1 Gpc). With a \emph{network} of detectors one could also measure distance. This opens up the possibility of studying the population evolution of black holes (and indirectly of the stars that produce them) \emph{over cosmological distances}.

\emph{Restricting component spins.} In the right panel of Fig.~\ref{f:compmasses_beta} we have plotted the error on the parameter $\beta$, which encodes the interaction between the components' spins and orbital angular momentum; its precise definition can be found in Ref.~3. To a first approximation one can neglect spin-induced precession of the orbital plane and take $\beta$ to be a constant. An important point is that if $|\beta| > 113/12$ then the spin of at least one component of the binary violates the Kerr bound, indicating a naked singularity, a boson star, or a still more exotic object. As seen in the right panel of Fig.~\ref{f:compmasses_beta}, EGO could measure $|\beta|$ to within 5\% of its abovementioned bound for masses up to almost $500\,M_\odot$, in stark contrast with Advanced LIGO. A more in-depth analysis has appeared elsewhere.\cite{amppapers} 

\emph{Other possible applications.} The large redshift reach of EGO would make it an ideal tool for cosmology; we confine ourselves to two more examples which were already foreseen by Schutz\cite{Schutz} in the context of LIGO and deserve to be revisited with a view on third-generation detectors. (i) With multiple detectors one can determine sky position and it becomes possible to identify the host galaxy (or cluster of galaxies), which will have some redshift $z$. From the gravitational wave signal the luminosity distance $D$ can be extracted. In a flat Universe there is a definite relationship $D(z)$ which depends on the Hubble constant $H_0$ as well as parameters $\Omega_0$ and $\Omega_\Lambda$ set by the mass density of the Universe and a possible cosmological constant, respectively.  Given a sufficient number of events at different distances one could fit the function $D(z)$, which would amount to measuring $H_0$, $\Omega_0$, and $\Omega_\Lambda$. (ii) At the largest scales, galaxy clusters tend to be on the surfaces of ``bubbles'' surrounding relative ``voids". It is natural to ask whether black hole binaries are similarly distributed, which may be relevant to dark matter studies.


\begin{thebibliography}{00}

\bibitem{Grishchuk} L.P.~Grishchuk, in \emph{Astrophysics Update}, ed.~J.W.~Mason (Springer-Praxis, Berlin, 2004).

\bibitem{Blanchet} L.~Blanchet, Liv.~Rev.~Rel.~{\bf 5}, 3 (2002).

\bibitem{amppapers} C.~Van Den Broeck and A.S.~Sengupta, Class.~Quantum Grav.~{\bf 24}, 155-176 (2007); C.~Van Den Broeck and A.S.~Sengupta, gr-qc/0610126.

\bibitem{Schutz} B.F.~Schutz, Class.~Quantum Grav.~{\bf 6}, 1761-1780 (1989).

\end{thebibliography}
\end{document}